\documentclass[twocolumn,showpacs,preprintnumbers,amsmath,amssymb,prl,superscriptaddress]{revtex4}
\usepackage{graphicx}

\begin{document}

\newcommand {\be}{\begin{equation}}
\newcommand {\ee}{\end{equation}}
\newcommand {\bea}{\begin{eqnarray}}
\newcommand {\eea}{\end{eqnarray}}
\newcommand {\nn}{\nonumber}
\newcommand {\ms}{m_{\rm s}^{\perp}}

\title{Bose-Glass Phases in Disordered Quantum Magnets} 

\author{Omid Nohadani}
\affiliation{Department of Physics and Astronomy, University of Southern California, Los Angeles, CA 90089-0484}

\author{Stefan Wessel}
\affiliation{Institut f\"ur Theoretische Physik III, Universit\"at Stuttgart, 70550 Stuttgart, Germany}

\author{Stephan Haas}
\affiliation{Department of Physics and Astronomy, University of Southern California, Los Angeles, CA 90089-0484}

\begin{abstract} 
In disordered spin systems with antiferromagnetic Heisenberg exchange, transitions into and out of a magnetic-field-induced ordered phase pass through a unique regime.
Using quantum Monte Carlo simulations to study the zero-temperature behavior, these intermediate regions are determined to be a Bose-Glass phase.
The localization of field-induced triplons causes a finite compressibility and hence glassiness in the disordered phase.
\end{abstract}


\maketitle

\noindent
Bose-Glass phenomena have recently been reported in a variety of disordered quantum many-body systems, including trapped atoms, vortex lattices, and Heisenberg antiferromagnets.
These experiments have in common signatures of finite compressibility in proximity to a Bose-Einstein condensate.
In atomic waveguides, the fragmentation of such a condensate  is caused by a random modulation of the local atomic density.\cite{frag}
It was shown that a quantum phase transition between the superfluid and the insulating Bose-Glass phase can be achieved under realistic experimental conditions.\cite{wang}
Furthermore, recent transport measurements of vortex dynamics in high-temperature superconducting cuprates have shown evidence of a Bose-Glass transition.\cite{vortex}
In the context of quantum antiferromagnetism, recent measurements of the magnetization and the specific heat have suggested a glassy regime in proximity to a magnetic-field-induced triplon condensate.\cite{tanaka_rand_bond,shindo_tanaka}
However, a theoretical understanding of the Bose-Glass phenomena in quantum spin systems based on a microscopic Hamiltonian is still lacking. 

In this letter, we study how disorder affects the quantum phase transition
between a valence bond solid and a magnetic-field-induced N\'eel-ordered
phase in an antiferromagnetic Heisenberg spin system.
Using large-scale quantum Monte Carlo
simulations down to ultra-low temperatures, we observe that in
cubic dimer systems with bond randomness, there is 
an intermediate Bose-Glass regime, separating an  
antiferromagnetically ordered phase of condensed triplons from a 
spin liquid phase of localized triplons at low magnetic fields. 
For weak inter-dimer
couplings, this model can be mapped onto a lattice boson model with
random potential.\cite{fisher89}
The N\'eel-ordered phase of delocalized triplons corresponds to the 
superfluid regime in the bosonic language. It is characterized by a
finite staggered magnetization perpendicular to the applied magnetic
field $\ms$, analogous to the superfluid order parameter in a bosonic system. 
The Bose-Glass phase is distinguished by a finite slope of the uniform magnetization $m_u$
as a function of the applied magnetic field, i.e., compressible bosons, 
whereas the order parameter $\ms$ vanishes.

In order to probe these observables and thus study emerging quantum phase transitions in such disordered quantum magnets, we apply the stochastic series expansion (SSE) quantum Monte Carlo (QMC) method.\cite{sandvik_green}
In particular, the directed-loop algorithm is used to minimize bounce probabilities in the 
loop construction when magnetic fields are applied to the system.\cite{directed_loop}
Ultra-low temperatures are chosen such that the relevant thermodynamic observables reflect true zero-temperature behavior.
In this work, we apply SSE QMC to a dimerized antiferromagnetic spin-1/2 Heisenberg model
on a cubic lattice, 
\bea
H = \sum_{\langle i,j \rangle} J_{ij} {\bf S}_i \cdot {\bf S }_j
- h \sum_i S^z_i,
\eea
where $J_{ij} = J$ for the intra-dimer couplings and $J_{ij} = J'$ for the inter-dimer couplings within and in between the planes.
$h$ denotes the external magnetic field.
Disorder is introduced by a bimodal distribution of the intra-dimer couplings, $P(J,x)=(1-x) \delta(J-J_1) + x\delta(J-J_2)$, with $J_2<J_1$  
and doping concentration $x$.
Typically, up to $500$ disorder realizations are included in the statistics for 
$L\le16$, and $8 \times 10^3/L$ for larger system sizes.

The order parameter $\ms$ can be calculated from the staggered structure factor,
\begin{equation}
\label{m_s}
S^\perp_{\rm s} = \frac{1}{L^3}\sum_{\langle i,j \rangle} (-1)^{i+j}
 \langle S^{x}_i S^{x}_j \rangle, \; \; \;\mbox{as}\; \;\;
m_{\rm s}^{\perp} = \sqrt{\frac{S^\perp_{\rm s}}{L^3}},
\end{equation}
where $L$ denotes the size of the system.
\begin{figure}[h]
\includegraphics[width=8cm]{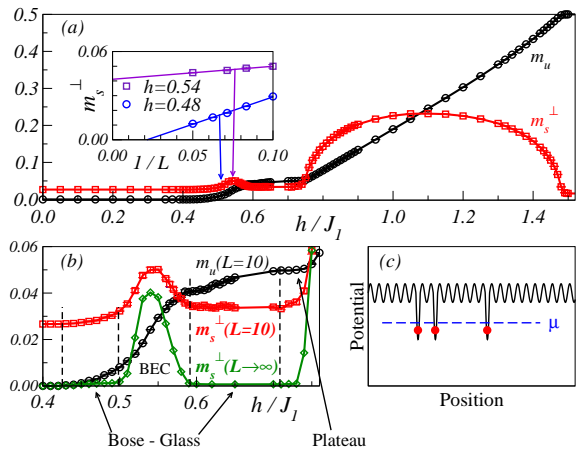}
\vspace{-2mm}
\caption{\label{bg}
(Color Online)
(a) Zero-temperature uniform and transverse staggered magnetization 
as a function of field for intra-dimer couplings
$J_1=2 J_2$, inter-dimer coupling $J'=0.1J_1$, and doping concentration $x=0.1$ for $10\times10\times10$ spins.
(b) magnification of the ``mini-condensation" surrounded
by two neighboring Bose-Glass phases, in which $m_u$
has a finite slope, whereas $\ms$ vanishes.
$m_u$ exhibits a plateau at the 
doping fraction of full polarization.
The quantum phase transition beyond the plateau is a Bose-Einstein 
condensation of triplons on the stronger dimer bonds.
The effective bosonic random-potential is illustrated in (c),
where the magnetic field corresponds to the chemical potential 
$\mu$, which controls the bosons (circles).
}
\end{figure}
Fig.~\ref{bg} shows $m_u$ and $\ms$ as a function of the applied 
magnetic field for $J_1=2J_2=10J'$ and $x=0.1$.
A rich field-dependence is observed.
For $h\ge0.75 J_1$, the singlet-triplet gap closes and $m_u$ increases with the magnetic field. The observed linear dependence on the applied field is expected from the XY universality class.\cite{pressure}
Furthermore, the external field induces a finite magnetic moment perpendicular to the field direction, $\ms$.\cite{rosch}
A square-root increase of $\ms$ is observed, indicating a field-induced Bose-Einstein condensation (BEC) of triplons, which extends up to the saturation field.
At higher fields ($h\ge1.5J_1$), all spins polarize fully along the field direction.
Zooming into the field region smaller than $0.75J_1$ reveals a small
bump in $\ms$.
For finite system sizes, $\ms$ is inversely proportional 
to the system length $L$ close to the lower critical field.\cite{pressure}
The inset of Fig.~\ref{bg}(a) shows $\ms$ for two field strengths, 
extrapolated to the thermodynamic limit.
While the offset vanishes for small fields, the small bump around $h=0.54J_1$ 
remains finite as $L\!\rightarrow\!\infty$, indicating an ordered phase of delocalized triplons from the doped bonds, which undergo a BEC.
Fig.~\ref{bg}(b) is a magnification of the interesting field region around this bump.
It also contains extrapolated data.
They reveal that for $0.44 J_1 \le h \le 0.5J_1$, the order parameter 
vanishes, whereas the uniform magnetization has a finite slope.
This signifies a new, disorder-induced phase prior to the BEC.

Since the inter-dimer coupling is chosen much smaller than both 
intra-dimer couplings,
this system can be mapped onto a hard-core boson model 
with a random potential as sketched in Fig.~\ref{bg}(c).
Deeper potential dips occur at random positions, reflecting that some of the intra-dimer couplings are weaker.
The chemical potential $\mu$, corresponding to the applied magnetic field, governs the occupation of hard-core 
bosons in the potential dips.
At small $\mu$, only the lowest minima are filled, and those spatially closer to each other cause islands of localized bosons.
The finite slope of the magnetization in Fig.~\ref{bg}(b), indicates a finite compressibility of the triplons in this picture.
Hence, in the region next to the BEC phase, the system is compressible, but not ordered because of triplon localization.
This is the manifestation of a Bose-Glass phase.
The more islands of compressible bosons are created, the 
larger the probability for the islands to come closer to each other.
Hence, there are enhanced correlations between 
the triplons due to the background interaction of the undoped bonds.
The localization disappears as soon as this interaction becomes relevant, 
which occurs at the BEC transition.
Therefore, each transition between the Mott-insulating and the superfluid 
phase should pass through a Bose-Glass regime.
\cite{fisher89}
This study delivers the first numerical evidence for the existence of a 
Bose-Glass phase in a microscopic spin model.

\begin{figure}[h]
\includegraphics[width=8cm]{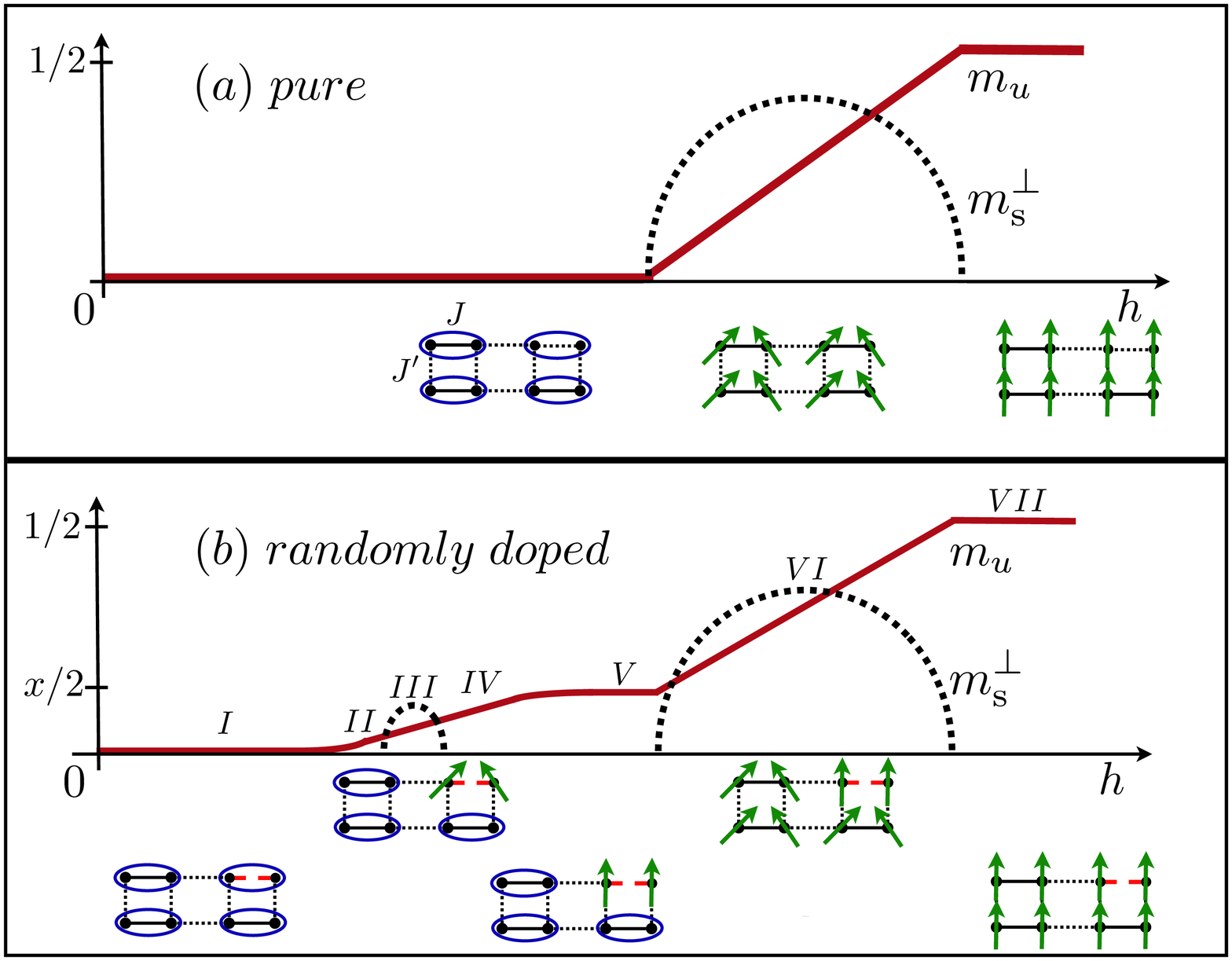}
%\vspace{-2.2cm}
\caption{\label{plateau}
(Color Online)
Schematic response of the zero-temperature uniform
and staggered magnetizations to an applied magnetic field.
Within the planes of the cubic lattice, dotted lines denote inter-dimer couplings $J'$ and solid lines the intra-dimer couplings.
At small fields, the weakly coupled dimers form a valence bond
solid state (elliptic bonds).
In the pure case (a), Bose-Einstein condensation occurs at the lower critical field.
At the saturation field, all spins are fully polarized, the system undergoes another BEC transition.
For the doped case (b),
intra-dimer bonds $J$ take the values $J = J_1$(solid lines)
or $J = J_2 $ (dashed lines).
A field scan reveals the following phases:
($I$) valence bond solid;
($II$) Bose-Glass phase;
($III$) ``mini-condensation";
($IV$) another Bose-Glass phase;
($V$) an intermediate plateau at $m_u= x \cdot m_u^{\mbox{sat}}$;
($VI$) BEC;
($VII$) full polarization.
}
\end{figure}

Fig.~\ref{plateau} provides a schematic picture of the different phases observed in the QMC data.
Planar sections of the cubic lattice are 
shown, containing weakly coupled spin dimers. 
In the clean case and
at sufficiently small fields, the dimer valence bond solid is
energetically the lowest state, as shown in Fig.~\ref{plateau}(a).
The quantum phase transition at the lower critical field may be regarded as a BEC of magnons in the lowest triplet branch.
Ultimately, at the upper critical field, all spins align fully along the field direction.
In the randomly doped case, Fig.~\ref{plateau}(b) shows seven possible phases.
The dimer valence bond solid is the ground state at small fields ($I$).
It requires a finite magnetic field strength to overcome the lowest singlet-triplet gap.
Since the doped bonds are weaker ($J_2<J_1$), these dimers break first.
Their spins respond to the increasing field, leading to a finite slope of uniform magnetization as a function of the applied field.
In the bosonic picture, this implies a finite compressibility of the field-induced triplons on the doped dimer-bonds.
These triplons are localized and the absence of phase coherence causes $\ms=0$.
Region ($II$) of Fig.~\ref{plateau}(b) illustrates this Bose-Glass phase.
Upon further increasing the magnetic field, delocalization of triplons sets in as they undergo a BEC transition, with $\ms>0$ as in region ($III$).
In Fig.~\ref{bg}(b), this phase occurs in the interval $0.5J_1\le h\le0.59J_1$, where the triplons on the doped bonds interact with each other via an exponentially small effective hopping term on the background 
of the remaining bonds ($J'_{\mbox eff}\ll J'$)~\cite{sigrist}, i.e. the triplons become delocalized.
In the bosonic picture, this ordered regime is the superfluid phase.
Upon further increasing the field, the spins align progressively along the field direction.
Eventually, $\ms$ vanishes and the delocalization disappears, which
constitutes another Bose-Glass phase upon exiting the ordered regime.
In Fig.~\ref{bg}(b), this occurs for $0.59J_1\le h\le0.71J_1$, corresponding to region ($IV$) of Fig.~\ref{plateau}(b).

The glassy phase of localized triplons disappears when all the spins of the doped bonds become fully polarized.
If the lower critical field of the undoped bonds $h_{c1}(J_1,J')$ is larger than the upper critical field of the doped bonds $h_{c2}(J_2,J'_{\mbox eff})$, a magnetization plateau is expected.
Region ($V$) of Fig.~\ref{plateau}(b) illustrates such a regime,
in which the uniform magnetization takes a constant value of $m_u=x \cdot m_u^{\mbox{sat}}$. Here, $m_u^{\mbox{sat}}$ is the
saturation magnetization and $x$ is the doping rate.
The present QMC data reveal a range of fields, for which such a plateau is observed, as shown Fig.~\ref{bg}(b).
Moreover, it is seen that a transition into and out of the superfluid phase passes through a Bose-Glass phase before entering the Mott-insulating phase, i.e. region ($III$) of Fig.~\ref{plateau}(b) is flanked by ($II$) and ($IV$) before entering ($I$) and ($V$), respectively.
Furthermore, there are no detectable  bond-disorder effects observed at and beyond the plateau, 
even though the fully polarized spins on the doped bonds are still randomly 
distributed in the system and should contribute to another Bose-Glass phase after the plateau.
This can be attributed to the negligible randomness effect at this level, 
since all of the spins on the doped bonds are saturated, both the hopping 
term $J'$ and the doping rate $x$ are small, and the doping obeys a bimodal distribution.

A further increase of the magnetic field breaks the remaining dimer singlets, 
as argued in section ($VI$) of Fig.~\ref{plateau}(b), thus driving the quantum phase transition 
to antiferromagnetic long-range order of delocalized triplons and
inducing a linear response to the magnetic field.
This transition is a field-induced BEC of triplons on the bonds with strong intra-dimer coupling $J_1$.
For $T>0$, the quantum critical field strength depends on temperature 
as $|h-h_c|\propto T_c^\alpha$, where in a narrow critical regime, 
$\alpha$ is determined to be $3/2$.\cite{universality}
This value agrees well with the mean-field prediction for BEC of bosons in the dilute limit.\cite{bec}
Ultimately, at very high fields, all spins align along the field direction,
and the system saturates magnetically, as illustrated in region ($VII$) of Fig.~\ref{plateau}(b).
At this threshold, another BEC with the same 
critical properties occurs. 
For this high-field transition
no Bose-Glass phases is detected, as argued previously.\cite{fss}
\begin{figure}[h]
\includegraphics[width=8cm]{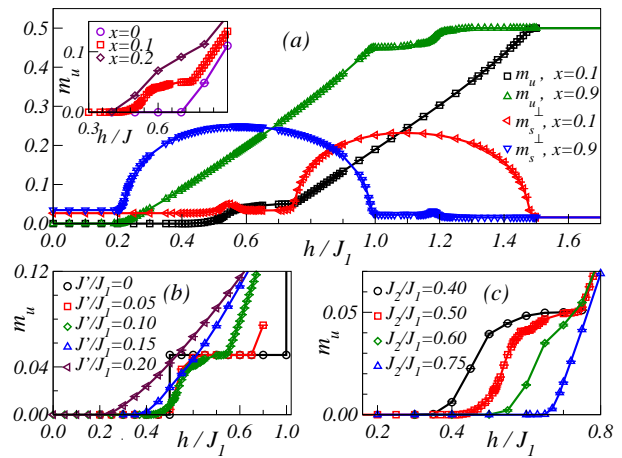}
\vspace{-2mm}
\caption{\label{umag}
(Color Online)
Zero-temperature uniform and staggered magnetization as a function of field
(a) for different doping concentrations $x$ at $J_1=2J_2=10J'$;
(b) for different inter-dimer couplings $J'$ between the decoupled
and strongly coupled dimer limits, with $J_1=2J_2$, and
$x=0.1$;
(c) for different intra-dimer coupling strengths of
the doped bonds $J_2$, with
$J_1=10J'$, and $x=0.1$.
}
\end{figure}

The dependence of $m_u$ and $\ms$ on doping concentration and the coupling strengths are studied in Fig.~\ref{umag}.
Different parameter sets are considered to explore the effects of bond
disorder close to the quantum phase transition.
The data for the doping rate $x=0.1$ in Fig.~\ref{umag}(a) are the same as shown in Fig.~\ref{bg}.
When $x=0.9$, analogous behavior is observed for the regime $0.98J_1\le h\le1.24J_1$, due to the abundance of weaker bonds $J_2$.
In this case, the effects of randomness, being two Bose-Glass phases flanking the superfluid phase, occur as a mirror image in the vicinity of the upper critical field $h=1.24J_1$ instead of the lower critical field $h=0.44J_1$.
A plateau with finite width occurs for $x=0.1$, as shown in the inset of Fig.~\ref{umag}(a).
For intermediate doping concentrations, $0.2<x<0.8$, the plateau is smeared out by the dimer-bond randomness.
Fig.~\ref{umag}(b) shows how the critical fields and the 
width of the plateau depend on the inter-dimer coupling $J'$.
The plateau has its maximum extent in the limit of decoupled dimers, i.e., $J'=0$.
This width decreases with increasing inter-dimer coupling strength and vanishes 
at a critical value $J'\approx 0.15 J_1$.
Therefore, simulations for $J'=0.1J_1$ reveal a finite width of the plateau as well as Bose-Glass phases flanking the triplon condensate on the weaker dimer bonds.
Furthermore, the ratio between the stronger and weaker intra-dimer bonds $J_1$ and $J_2$ controls the width of the plateau, as shown in Fig.~\ref{umag}(c).
If the values of $J_1$ and $J_2$ are too close, the effects of 
randomness are suppressed, smearing out both the magnetization plateau and the Bose-Glass phase.
However, upon decreasing the ratio $J_2/J_1$, a magnetization plateau appears. 
A ratio of $J_2/J_1=1/2$ was found to be sufficiently low to clearly reveal the novel, disorder-induced quantum phases.

Indications of a Bose-Glass phase between a gapped incompressible 
phase and a field-induced antiferromagnetic phase were recently 
suggested by high-field magnetization measurements on bond-disordered 
$\rm Tl_{(1-x)}K_xCuCl_3$ for $x<0.36$.\cite{tanaka_rand_bond}
More recent specific heat measurements on this compound with doping 
rates up to $x\le0.22$ examined the effect of randomness on the phase 
boundaries as a function of temperature. They
observed the emergence of a novel phase prior to the field-induced BEC.\cite{shindo_tanaka}
However, the linear response of the measured magnetization 
to the applied field starting at $h=0$ indicates that non-magnetic 
$\rm K$-doping of $\rm TlCuCl_3$ not only introduces bond-disorder, but also 
a pronounced directional Dzyaloshinskii--Moriya vector.\cite{dm}
Therefore, Bose-Glass effects are likely to be suppressed.
Hence, doped compounds with negligible spin-orbit coupling 
and vanishing directionality are expected to reveal Bose-Glass features.

\begin{figure}[h]
\includegraphics[width=8cm]{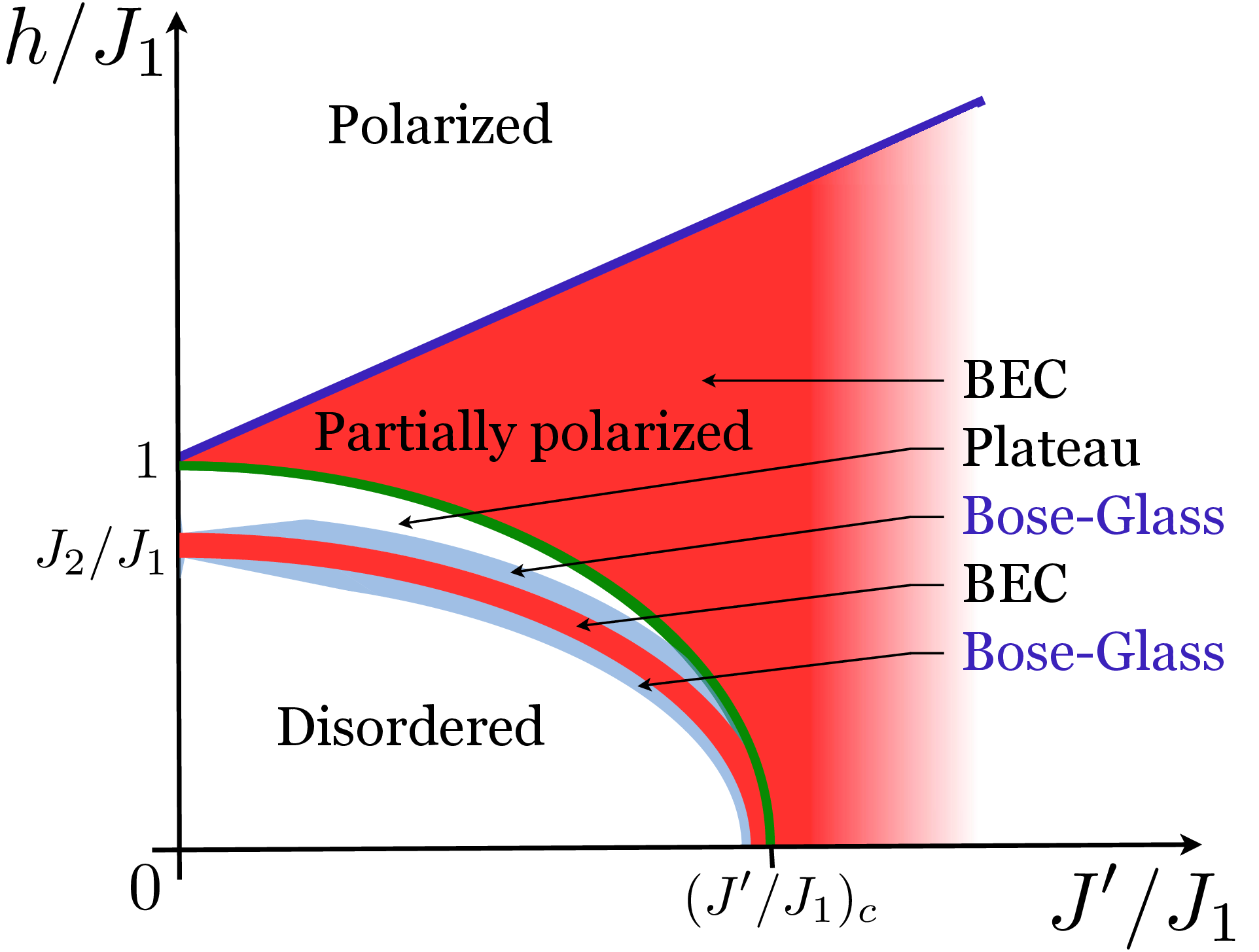}
\vspace{-1mm}
\caption{\label{phase}
(Color Online)
Zero-temperature phase diagram of three-dimensional weakly coupled dimers with random 
intra-dimer coupling at a doping rate of $x\le15\%$.
The plateau is most pronounced at weak inter-dimer couplings.
For $(J'/J_1)>(J'/J_1)_c\approx0.249$, the gap vanishes,
and the order sets in at infinitesimal fields.
For small $x$, we do not expect to detect any effects of randomness 
at saturation fields.
}
\end{figure}

We conclude by proposing a phase diagram of weakly coupled dimers with 
random intra-dimer coupling strengths ($J_1>J_2$) in Fig.~\ref{phase}.
Quantum Monte Carlo data show that at finite randomness, a field-induced quantum phase transition into and out of an ordered Bose-Einstein condensate passes through a Bose-Glass phase.
The localization of the bosons and the finite compressibility manifests this unique regime.
Once delocalized, the triplons condense and N\'eel-order sets in.
Depending on coupling ratios, an intermediate plateau can occur, in which the spins of the doped bonds are fully polarized.
This rich field-dependence is expected to be experimentally detectable  in weakly coupled dimer compounds with small doping and negligible spin-orbit coupling or directionality effects. 

We thank T. Roscilde, P. Schmidt, M. Troyer, and T. Vojta for useful  discussions. 
Furthermore, we acknowledge financial support from NSF Grant No. 
DMR-0089882. SH and SW appreciate the hospitality of the Kavli Institute for Theoretical Physics.
Computational support was provided by the USC Center for High Performance Computing and Communications.


\begin{thebibliography}{99}
\bibitem{fss}
For $h\ge1.5J_1$, finite-size effects are responsible for the offsets in $\ms$, which scales as $\ms=1/\sqrt{2L^3}$.\cite{hc2}
\end{thebibliography}
\end{document}